\documentclass[referee]{raa}
\usepackage{graphicx,times}  
\usepackage{natbib}
\usepackage{amssymb,amsmath}
\bibpunct{(}{)}{;}{a}{}{,}

\newcommand{\NVI}{\mbox{{N\,{\scriptsize VI}}}}
\newcommand{\NVII}{\mbox{{N\,{\scriptsize VII}}}}

\newcommand{\OVII}{\mbox{{O\,{\scriptsize VII}}}}
\newcommand{\OVIII}{\mbox{{O\,\scriptsize VIII}}}

\newcommand{\NeX}{\mbox{{Ne\,{\scriptsize X}}}}

\newcommand{\FeXVII}{\mbox{{Fe\,{\scriptsize XVII}}}}

\newcommand{\Lya}{\ensuremath{\hbox{Ly}\alpha~}}

\newcommand{\A}{\AA~}
\def\xmm{{\it XMM-Newton }}
\def\cha{{\it Chandra }}

\newcommand{\Os}{\mbox{{O$^{5+}$}}}
\newcommand{\Ov}{\mbox{{O$^{6+}$}}}

\begin{document}

\title{On the diffuse soft X-ray emission from the nuclear region of M51} 

   \volnopage{Vol.0 (200x) No.0, 000--000}      
   \setcounter{page}{1}          

   \author{Jiren Liu
      \inst{1}
   \and Shude Mao
      \inst{2,1,3}
   }

 \institute{National Astronomical Observatories, Beijing 100012, China; {\it jirenliu@bao.ac.cn}\\
        \and
Physics department and Tsinghua Center for Astrophysics, Tsinghua University, Beijing, 100084, China \\
		\and
		  Jodrell Bank Centre for Astrophysics, University of Manchester, Manchester, M13 9PL, UK \\
}

\date{Received~~2009 month day; accepted~~2009~~month day}

\abstract{
We present an analysis of the diffuse soft X-ray emission from the nuclear region of M51
combining both \xmm RGS and \cha data. Most of the RGS spectrum of M51 can be fitted 
with a thermal model with a temperature of $\sim0.5$ keV
except for the \OVII\ triplet, which is forbidden-line dominated.
The Fe L-shell lines peak around the southern cloud,
where the \OVIII\ and \NVII\ \Lya lines also peak.
In contrast, the peak of the \OVII\ forbidden line is about 10$''$ offset from 
that of the other lines, indicating that it is from a spatially distinct component.
The spatial distribution of the \OVII\ triplet mapped by the \cha data
shows that most of the \OVII\ triplet flux is located at faint regions near edges, 
instead of the southern cloud where other lines peak. 
This distribution of the \OVII\ triplet is inconsistent with the photoionization model.
Other mechanisms that could produce the anomalous \OVII\ triplet, including 
a recombining plasma and charge exchange X-ray emission, are discussed.
}

	\authorrunning{J. Liu \& S. Mao}
   \titlerunning{Diffuse soft X-ray emission of M51}

	\maketitle

\section{Introduction}

The Whirlpool galaxy M51 (NGC 5194) is classified as a LINER or Seyfert 2 galaxy
from studies of optical emission lines \citep[e.g.,][]{Sta82,Ho97}. 
The close distance of M51 (7.8 Mpc, the mean 
value taken from NED\footnote{http://ned.ipac.caltech.edu} database) allows detailed studies 
of its nuclear activity. 
Radio observations show a bipolar outflow, comprising a southern cloud and a northern loop, 
which is also seen in the optical emission line map \citep{For85,Cra92}.
The optical outflow with velocities 
as high as 1500 km s$^{-1}$ has been reported by \citet{Cec88}, who also found that 
the arcuate radio emission of the southern
cloud lies inside the bright optical emission line region, indicating a bow shock caused
by the radio jet emanating from the nucleus. 

The interaction between the radio jet/outflow and the surrounding gas represents an important 
feedback mode of active galactic nucleus (AGN), which is crucial for our understanding of galaxy
evolution \citep[e.g.][]{Fab12}. X-ray observations are very effective in studying such 
phenomena as the radio jet-driven outflow will heat the surrounding gas to X-ray 
emitting temperatures. Extended X-ray emission from M51 was first detected using the
{\it Einstein Observatory} \citep{Pal85},
and then with {\it ROSAT} \citep{Mar95,Ehl95}. The {\it BeppoSAX} data 
of M51 showed that the X-ray emission of the nucleus is only seen directly above 10 keV,
implying a neutral hydrogen column density of 
$N_{\rm H} \sim10^{24}$ cm$^{-2}$ \citep{Fuk01}.

Observations of {\it Einstein} and {\it ROSAT} are limited by their spatial resolution. 
With its sub-arcsec angular resolution, \cha has provided new insights on M51.
\citet{TW01} found that the nuclear X-ray morphology of M51 is similar to that seen 
in radio and optical observations. The X-ray spectra of the southern cloud and northern loop
are similar and can be fitted by a thermal model of temperature $\sim0.55$ keV.
The X-ray morphology and spectra suggest that the nuclear X-ray emitting gas of M51 
are shock-heated by the bipolar radio outflow from the nucleus.
Both the X-ray point sources and 
the extended X-ray emission from the disk of M51 have been studied 
using \cha \citep[e.g.,][]{TW04,Tyl04} and \xmm data
\citep[e.g.,][]{Dew05,OW09}.

All these X-ray studies are based on CCD data, for which the  
spectral resolution is around 10 at 1 keV. In contrast, 
the Reflection Grating Spectrometers (RGS) aboard the \xmm telescope \citep{den01}
have a better spectral resolution for moderately 
extended sources due to their large dispersion power. 
The RGS spectral resolution is $\sim0.14\theta$\,\AA, where $\theta$ (in units of arcmin) 
is the angular extent of the source along the dispersion direction. 
For M51, the spatial extent of the nuclear X-ray emitting region is about
0.5$'$, which corresponds to  
a spectral resolving power $\lambda/\delta_\lambda\sim150$ at 15 \AA\ 
given the spatial resolution of {\it XMM-Newton}.
With this resolution many emission lines, especially the He-like \OVII\ triplet
which is a diagnostic tool for the thermal state of plasma, are well resolved.

The \OVII\ triplet consists of a resonance line, two inter-combination lines and 
a forbidden line. For an optically-thin thermal plasma in ionization equilibrium, the electron 
collisional excitation is efficient and favors the resonance line.
In \citet{Liu12}, we studied nine nearby star-forming galaxies (including M51) 
and found that the forbidden lines of their \OVII\ triplets are comparable to or even stronger 
than the resonance lines. 
We proposed the charge-exchange process between highly ionized ions and neutral
species as a possible explanation. The charge-exchange captured electrons of the recipient ions are in  
excited states and their downward cascading favors the forbidden line \citep[e.g.][]{Den10}.
In the special case of M51, however, the explanation is complicated by the presence 
of the central low-luminosity AGN and the radio jet-driven outflow.

In \citet{Liu12}, only the line ratio of the \OVII\ triplet of M51 was studied.
In this paper we provide a more detailed study of the \xmm RGS spectrum of M51.
In particular, we study the spatial distribution of emission lines
along the cross-dispersion direction of RGS. As shown in \S 3,
the spatial distribution of the \OVII\ triplet is different from other lines. 
Given this situation, the \cha image of M51 with sub-arcsec angular resolution is helpful
in revealing the \OVII\ triplet distribution.
Thus we also analyze the archival \cha data of M51 with an accumulated
exposure time of 700 ks, which is much deeper than the short exposure data ($\sim15$ ks) 
used in previous studies \citep[e.g.][]{TW01}.

We describe the observational data in \S\ 2 and present the results
of \xmm  and \cha data in \S\ 3 and \S\ 4, respectively. Discussion of the results 
is covered in \S\ 5.
Throughout the paper, the errors quoted are for the 90\% confidence level. 
At a distance of 7.8 Mpc, 1 arcsec corresponds to 38 parsec.

\section{Observational data }

The two RGSs on-board the \xmm telescope are slit-less dispersive spectrometers, and 
photons from extended sources are recorded on CCD detectors with the 
dispersion angle and 1D spatial information along the cross-dispersion direction. 
Because the dispersion directions are different for different observations,
to ensure that the dispersed spectra are from similar spatial regions, 
we use four archival datasets of \xmm RGS observations of M51 as listed in Table 1.
The maximum difference of position angles between them is about 30 degrees.
The total effective exposure time is $\sim$ 100 ks after 
removing intense flare periods. 

\begin{table}
	\caption{List of \xmm RGS observations of M51}
	\begin{tabular}{ccccc}
		\hline
		 ObsID &  $t_{\rm tot}$ (ks)& $t_{\rm eff}$ (ks)& Obs time & P.A. (degrees)\\
		\hline
		 0212480801 &  49 & 26& 2005-07-01 & 294\\
		 0303420101 &  54 & 35& 2006-05-20 & 326\\
		 0303420201 &  37 & 25& 2006-05-24 & 323\\
		 0677980701 &  13 & 13& 2011-06-07 & 312\\
		\hline
	\end{tabular}
	\begin{description}
			\begin{footnotesize}
			\item
				Note: $t_{\rm tot}$ is the total exposure, $t_{\rm eff}$ is
				the useful exposure after removing periods of flares, and P.A.
				is the position angle.
			\end{footnotesize}
	\end{description}
\end{table}

The most recent version of the Science Analysis System (SAS 14.0) of \xmm is used for
the reduction of photon events. 
As stated above, the RGS spectrum is broadened by the
spatial extent of the source along the dispersion direction. To produce a broadened
redistribution matrix file (RMF), we convolve the RMF produced by the SAS tool {\it rgsproc}
with a \cha image of M51 (0.4-2 keV) using the {\it rgsrmfsmooth} tool written by Andrew Rasmussen.
The broadened RMF is calculated separately for each dataset.
For background subtraction, we use the model background generated by 
{\it rgsproc} based on the flux of CCD 9 beyond 1$'$ from the on-axis position.
Because M51 lies at $\sim1'$ away from the on-axis position for most 
observations listed in Table 1, the model background is over-predicted.
Indeed, the model background exceeds the observed spectrum at $\lambda>28$ \AA. 
Thus we scale down the model background by a factor of 0.6 based on 
the event distribution of CCD9.
Since we focus on emission lines, the uncertainty in the background subtraction 
will not affect our major results.

\begin{figure}
\includegraphics[height=2.5in]{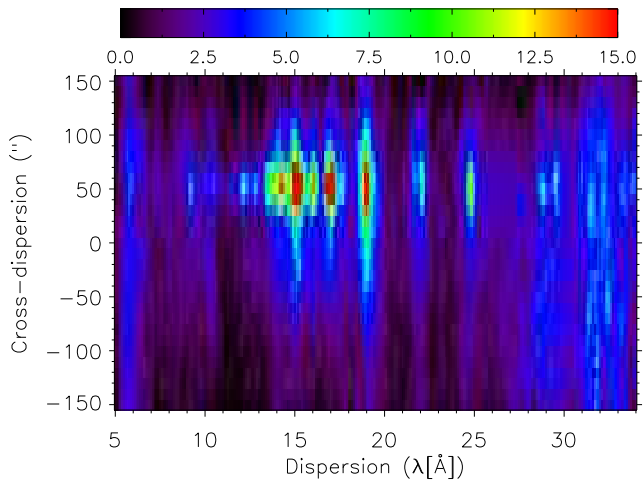}
\caption{Cross-dispersion vs dispersion CCD image of M51 combining both
RGS1 and RGS2 data of all four observations.
The emission lines are seen as vertical contours and are well resolved.
The color bar is for the total event counts on linear scale.
The zero point along the cross-dispersion direction of 
ObsID 0303420201 is adopted. 
}
\end{figure}

\begin{figure*}
\includegraphics[height=2.8in]{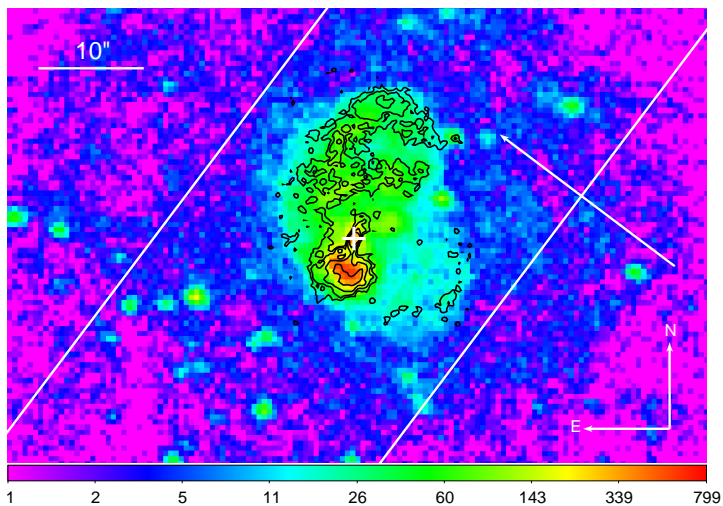}
\caption{\cha counts image of M51 within $0.4-2$ keV (logarithmically scaled). 
The overlaid radio contours (1.5, 3, 6 and 12 in units of 10 $\mu$Jy) are from VLA observations 
at 6 cm wavelength by \citet{Cra92}.
The two solid lines separated by 30$''$ indicate the RGS dispersion direction of ObsID 0303420201, 
while the arrow indicates the positive cross-dispersion direction.
The white cross marks the position of the nucleus of M51, which corresponds to a cross-dispersion
distance of 52$''$ for ObsID 0303420201.
}
\end{figure*}

Because CCD 4 of RGS2 covering $20-24$ \A failed early in
the mission, the data of the \OVII\ triplet around 22 \A are only from RGS1.
Similarly, CCD 7 of RGS1 also failed and the data within $10.6-13.8$ \A 
of RGS1 are missing. Figure 1 presents the CCD image of the \xmm RGS data 
of M51 combining both RGS1 and RGS2 for all four observations. 
The offsets between RGS1 and RGS2 and between different observations have
been corrected. The figure shows clearly the well-resolved emission lines. 

We use seven \cha datasets with ObsID numbers of 13812, 13813, 13814, 13815, 13816, 15496, 
15553, as observed by PI Kip Kuntz.
After removing the flare periods, the total effective exposure time is about 700 ks.
The datasets are analyzed with {\it CIAO} (version 4.6) following standard procedures.
As an illustration, Figure 2 presents the counts image of M51 within 0.4-2 keV created by merging 
all the seven datasets. The over-plotted contours are from the radio 6 cm observation \citep{Cra92},
and show a close correspondence with the X-ray image. The \xmm RGS dispersion 
direction is also illustrated in Figure 2.

\section{XMM-Newton RGS Results}

\begin{figure}
\includegraphics[height=2.1in]{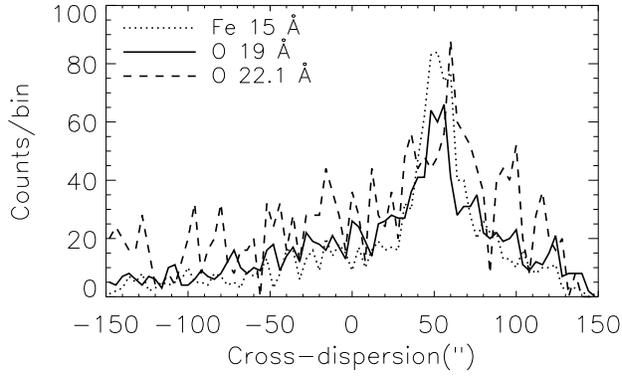}
\caption{Spatial distribution of the emission lines of \FeXVII\ at 15 \AA, \OVIII\ \Lya at 19 \AA,
	and \OVII\ forbidden line at 22.1 \AA.
For clarity, the profile of the \OVII\ forbidden line at 22.1 \AA\ has been multiplied by a factor of 4.
}
\end{figure}

\begin{figure}
\includegraphics[height=2.2in]{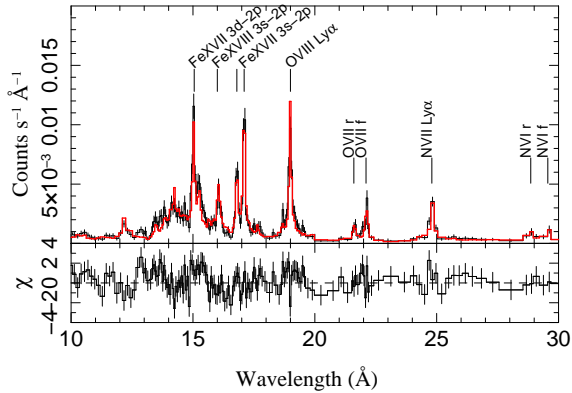}
\caption{Background-subtracted \xmm RGS spectrum of M51 combining both RGS1 and RGS2 data of
all four observations. Red histograms are the fitted thermal model with four Gaussians. 
$\chi$ is the difference between data and model divided by the error.
}
\end{figure}

\subsection{Spatial distribution of X-ray emission lines }

First we study the spatial distribution of the well-resolved X-ray emission lines
along the cross-dispersion direction of \xmm RGS.  
The \FeXVII\ lines at 15 and 17 \AA, the \OVIII\ and \NVII\ \Lya lines at 19 and 24.8 \AA, and 
the \OVII\ forbidden line at 22.1 \A are studied. 
The wavelength regions adopted to calculate the photon counts are
14.8 - 15.5 \AA, 16.6 - 17.4 \AA, 18.6 - 19.4 \AA, 24.4 - 25.2 \AA, and 21.8 - 22.4 \AA,
respectively. We find that the profiles of \FeXVII\ lines at 15 and 17 \AA\ are similar,
as are the \OVIII\ and \NVII\ \Lya lines. For clarity, only the profiles of \FeXVII\ line at 15 \AA, 
\OVIII\ \Lya line at 19 \AA, and \OVII\ line at 22.1 \A are plotted in Figure 3. 
We see that the bright \FeXVII\ line at 15 \A is centrally peaked around 
50$''$, with a FWHM (full width at half maximum) of $\sim20''$. This FWHM
is similar to the spatial resolution of \xmm RGS. Since the cross-dispersion position of 50$''$
corresponds to the location of the southern cloud, it means that
the emission lines of \FeXVII\ are mainly from the compact southern cloud.

The \OVIII\ \Lya line at 19 \A is also peaked around 50$''$, the same as the \FeXVII\ lines.
The spatial distribution of the \OVII\ forbidden line at 22.1 \AA, however, is different 
from those of the other lines.
Its maximum is around 60$''$, 10$''$ offset from the peaks of other lines. 
This indicates that the \OVII\ forbidden line 
is from a spatially distinct component, compared with the other lines. 

\begin{table}
\caption{Fitting results}
 \tabcolsep 2pt
\begin{tabular}{ccccccccccc}
\hline
T (keV) & N & O& Ne& Fe & \OVII$_r$ & \OVII$_f$ & \NVI$_r$ & \NVI$_f$ & $\chi^2_{\nu}$ \\
\hline
 $0.52\pm0.02$ & $4.4\pm1.1$ & $0.6\pm0.1$ & $0.6\pm0.2$ &
 $0.3\pm0.1$&$1.3\pm0.7$&$4.1\pm1.0$&$2.2\pm0.9$&$2.5\pm1.2$&0.91 \\
\hline
\end{tabular}
\begin{description}
\begin{footnotesize}
\item
	Note:	
	the abundances are relative to the solar values; \OVII$_{r,f}$ and \NVI$_{r,f}$ (in units of
	10$^{-5}$ photons s$^{-1}$ cm$^{-2}$) are the line intensities
of \OVII\ resonance line, \OVII\ forbidden line, \NVI\ resonance line, and \NVI\ forbidden line,
respectively. 
\end{footnotesize}
\end{description}
\end{table}

\subsection{X-ray spectrum of M51}

Since the X-ray emission lines have different spatial distributions, it is optimal 
to extract spectra within different spatial regions. However, the difference 
of the scale of $\sim10''$ is too small to allow such spectral extractions from \xmm RGS.
Thus we only extract one spectrum centered on the nucleus of M51 with a cross-dispersion
width about 1$'$.  
The background-subtracted spectrum of M51 combined from both RGS1 and RGS2 data of
all four observations is plotted in Figure 4. 

We see that the spectrum of M51 is dominated by emission lines. 
The \OVII\ triplet is dominated by the forbidden line at 22.1 \AA\ and
the \NVI\ triplet also has a forbidden line comparable with its resonance line.
The bright Fe L-shell lines around 15 - 17 \A indicate that they are not from a 
photoionized plasma, for which the Fe L-shell lines are expected to be much weaker
\citep[e.g.,][]{Kal96,Sak00}.
They are likely due to a collisionally ionized plasma as suggested in previous 
studies \citep[e.g.,][]{TW01}.
Thus we fit an optically thin collisional-ionization-equilibrium (CIE) thermal model
\citep[vapec,][]{Fos12} to the observed RGS
spectrum of M51. Only the abundances of the elements of Ne, Fe, O, and N, which show 
bright emission lines in the spectrum of M51, are allowed to vary. 
The abundances of other elements are set to solar values \citep{Lod03}.

On the other hand, the forbidden-line dominated \OVII\ triplet of M51 shows a spatial 
distribution different from other lines.
As stated in \S\ 1, it is impossible to explain such an anomalous triplet 
by a thermal CIE model.
Thus we add two Gaussians to represent the \OVII\ resonance and forbidden lines.
The inter-combination line is too weak to be fitted. Similarly, we add another two Gaussians
to represent the \NVI\ resonance and forbidden lines. 
The model is subject to an absorption model of {\it wabs} \citep{MM83}
with a foreground neutral hydrogen column density of $2 \times 10^{20} \rm cm^{-2}$ \citep{Nh05}.
The fitted results are listed in Table 2 and over-plotted in Figure 4.

The model provides a reasonable fit to the observed spectrum of M51. 
The fitted temperature is 0.52 keV, similar to that obtained by \citet{TW01}. However, 
the fitted abundances are higher than their values,
which are only around 0.1 solar value. The low abundances 
they obtained are most likely due to the poor spectral resolution of \cha data. 
We note that the fitted O and N abundances are not affected by the Gaussian 
fitting to the corresponding \OVII\ and \NVI\ triplets, since they are most likely
due to a mechanism different from CIE models and at the fitted temperature, 
the emissivity of \OVII\ and \NVI\ triplets of the CIE model is weak.
The residuals around the Fe and O lines are likely due to the simple convolution 
modeling of the RMF.
There are residuals around 13.5 \AA, indicating that the \NeX\ triplet is under-estimated.
The residuals around 18.6 \A are likely due to the \OVII\ He$\beta$ line. 

The temperature diagnostic G ratio of the He-like triplet for a thermal CIE plasma is defined
as \citep{GJ69}
\begin{equation}
{\rm G}=\frac{{\rm f}+{\rm i}}{{\rm r}},
\end{equation}
where f, i, and r represents the intensity of the forbidden, inter-combination, 
and resonance lines, respectively.
If we assume the intensity of the 
inter-combination line is 1/4 of that of the forbidden line \citep{Smi01},
the fitted G ratio of the \OVII\ triplet is $3.9\pm2.3$.
Similarly, the fitted G ratio of the \NVI\ triplet is $1.4\pm0.8$.
The G ratio of \OVII\ triplet shows that the \OVII\ triplet is not due to 
 CIE thermal plasmas, for which the G ratios are generally smaller
than 1. The G ratio of \NVI\ triplet is barely consistent with a CIE model around 10$^6$ K.

\begin{figure*}
\includegraphics[height=2.6in]{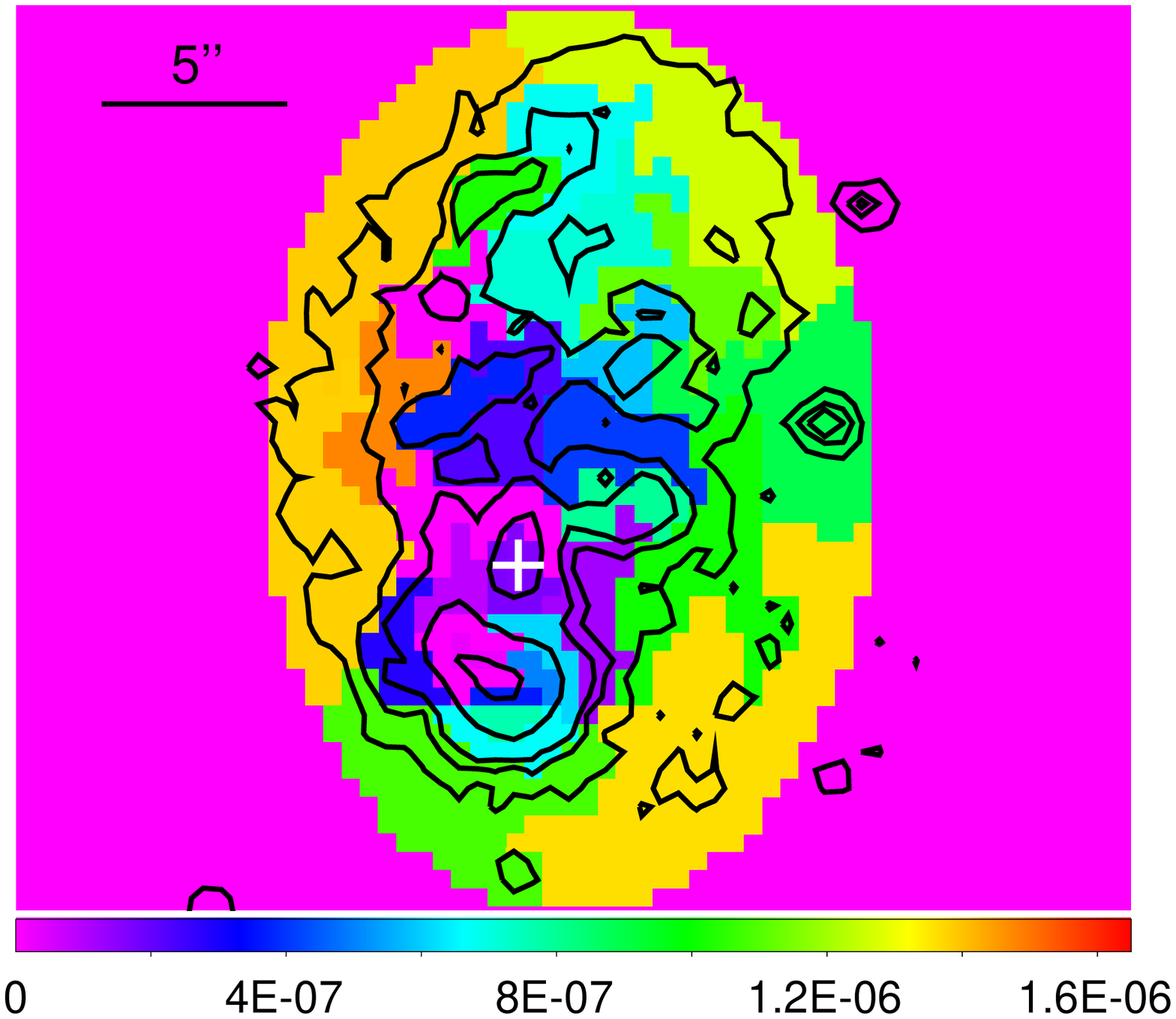}
\includegraphics[height=2.6in]{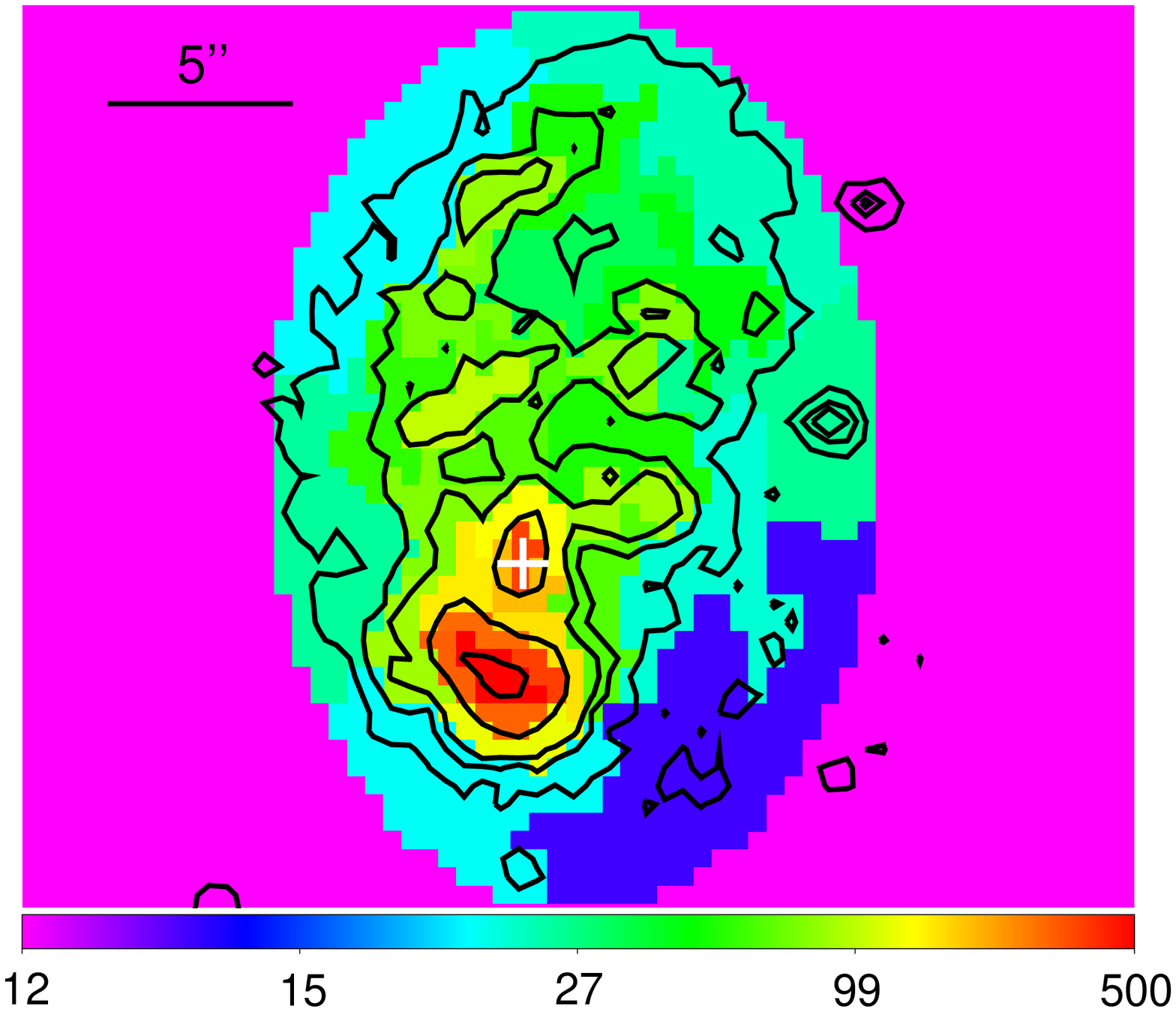}
\caption{Left: Fluxes of the \OVII\ triplet obtained with the \cha data of M51 using 
the contour binning method. Right: The binned counts map of M51 within 0.4 - 2 keV.
The contours of levels of 20, 50, 80, 200, and 550 for unbinned map are plotted
in both panels. The white cross marks the nucleus of M51. 
}
\end{figure*}

\section{\cha results}

As shown in \S 3.1, the peak of the spatial profile of the \OVII\ forbidden line is offset
from those of the other lines by $\sim10''$. This scale can be easily resolved with \cha data.
Thus we analyze the \cha data of M51 in this section. 

We use the contour binning method \citep{San06} to divide the nuclear X-ray emitting 
region of M51 into 34 different bins with a minimum signal to noise ratio of 40.
The X-ray spectrum is extracted from each bin with
a background spectrum taken from a source-free region outside the nuclear region. 
We fit a thermal CIE model with variable O, Ne, Mg, and Fe abundances
and limit the fitting range to $0.4 - 1.5$ keV.
We include a Gaussian (with a center of 21.85 \A and a width of 0.25 \AA) to 
represent the \OVII\ triplet. Adding a Gaussian 
will improve the $\chi^2$ by 4.3 on average. The improvement of $\chi^2$ is
8.1 for the spectra with the Gaussian line fluxes larger than 
$5\times10^{-7}$ photons s$^{-1}$ cm$^{-2}$. 
We also tested the fitting method by fitting mock spectra with a line flux of 
$1\times10^{-6}$ photons s$^{-1}$ cm$^{-2}$, and found that about 75\% of the fitted 
fluxes are within 3 sigma of the true value. 

Due to the limited spectral resolution of \cha data, we do not expect the fitted
\OVII\ triplet fluxes to be as accurate as the measurement by the \xmm RGS data. 
Nevertheless, it is likely to provide a qualitative distribution of the \OVII\ triplet,
especially for bins with high \OVII\ triplet fluxes.
The fitted fluxes of the \OVII\ triplet from the \cha data are plotted
in the left panel of Figure 5. For comparison, the binned counts image 
is plotted in the right panel. 
We see that the fluxes of the \OVII\ triplet are generally higher for 
faint bins near the edges. It shows that the fitted \OVII\ triplet has a more extended 
distribution than that of the total emission within $0.4-2$ keV, which is centered on the 
southern cloud. The northern regions have more \OVII\ triplet flux than the southern part.
This is consistent with the \xmm result. 
Another interesting feature to note is that the southern cloud 
seems to be enclosed by an arc of \OVII\ triplet in the outward direction.

\section{Discussion}

We have analyzed the \xmm RGS spectrum of M51, the emission lines of
which are well resolved. Most of the spectrum of M51 can be fitted with 
a thermal CIE model of temperature $\sim0.5$ keV, except for the \OVII\ triplet, 
which is forbidden-line dominated. The \FeXVII\ lines at 15 and 17 \A 
are centrally peaked around the cross-dispersion position of 50$''$ with a 
FWHM of 20$''$. This indicates that they are mainly from 
the compact southern cloud. The \OVIII\ \Lya line at 19 \A and \NVII\ \Lya line
at 25 \A also peak around the same position as the \FeXVII\ lines. In contrast,
the peak of the \OVII\ forbidden line is about 10$''$ offset from the other lines. 
The \OVII\ triplet map obtained with the \cha data
shows that most of the fluxes of the \OVII\ triplet are located at faint regions,
instead of the southern cloud where the other lines peak around.

As stated in \S\ 1, the forbidden-line-dominated \OVII\ triplet of M51 cannot be due to 
a thermal CIE plasma, and we have proposed the charge-exchange process as a possible explanation.
Nevertheless, the existence of low-luminosity AGN and the 
radio jet-driven outflow in the nuclear region of M51 makes other explanations possible,
including photoionization and non-equilibrium-ionization plasmas.
Below we discuss them in turn.

For a photoionized plasma, the emission is dominated by recombination, which 
also favors the forbidden line.
One characteristic feature of the photoionization model is the spatial profile. 
As the central AGN is the ionizing 
source, the ionizing flux will decline as $r^{-2}$ and photoionization is most important for 
regions close to AGN. This has been illustrated by the optical study of M51 \citep{Bra04}, which
shows that photoionization is dominant within the inner region ($r<1''$) and 
the shock model is preferred outside ($r\sim2.5''$). 
The mapped morphology of the \OVII\ triplet from \cha data shows no fluxes around the nucleus and 
is most prominent at the outer regions. This is inconsistent with the photoionization model.

All the optical, radio and X-ray studies of M51 support a scenario in which
the nuclear gas of M51 is shock-heated by the radio outflow emanating from the nucleus
\citep{Cec88,For85,TW01}. In this situation, the non-equilibrium-ionization (NEI) plasma is 
another possible explanation for the anomalous \OVII\ triplet.
If a plasma is shock-heated suddenly, the ionization process lags behind heating, and 
will result in an ionizing plasma.
The inner-shell collisional ionization of \Os\ ions can lead to excited
\Ov\ ions and enhance the forbidden line emission \citep[e.g.,][]{L99}.
However, this only happens in a short transition period when the \Os\ fraction is
non-negligible. 
Using the {\it Sedov} model \citep{Bor01}, we find that the 
G ratio of the \OVII\ triplet is around 3 at an ionization time 
$n_et\sim3.5\times10^9$ cm$^{-3}$s. This timescale is too short compared 
with the spatial extent we studied. A more likely explanation is a recombining NEI plasma.

When a blast wave expands into a rarefied medium, a recombining plasma could be produced due to 
rapid adiabatic cooling, as
proposed to explain the radiative recombination continuum features in some supernova  
remnants (SNRs) \citep[e.g.,][]{Yam12}. The spatial distribution of the \OVII\ triplet 
mapped by the \cha data is consistent with this scenario. 
A unique test of a recombining plasma is the associated radiative recombination continuum.
However, due to the presence of strong Fe lines, 
it is hard to tell whether there is a radiative recombination continuum of \OVII\ (at 16.8 \AA) 
in the spectrum of M51. 

On the other hand, to explain the anomalous \OVII\ triplet with the charge-exchange process, the 
interacting interfaces between the highly ionized gas and neutral species are needed.
The observed HI \citep{Wal08} and CO \citep{Kod11} maps of M51 are well correlated with 
the spirals. CO is also detected on 1-2 arcsec scales around the nucleus \citep{Mat07}.
Currently there is no much information of neutral materials on 10$''$ scales of M51. 
Further observations are needed to test the existence of neutral materials on 10$''$ scale.
Different from the recombining plasma, 
the charge-exchange model shows no radiative recombination continuum, but 
has enhanced emission lines from preferred energy levels ($n=4-7$ for OVII) 
\citep[e.g.][]{Bei03}. 
Future spatially-resolved high-resolution X-ray spectroscopy by the soft
X-ray spectrometer of Astro-H 
will be able to test the existence of the radiative recombination continuum of M51.

\begin{acknowledgements}
We thank the referee for their valuable comments and Lijun Gou and Richard Long 
for reading of the draft.
This work is supported by a National Natural Science Foundation of China for
Young Scholar Grant (11203032), and by the Strategic Priority
Research Program ``The Emergence of Cosmological
Structures`` of the Chinese Academy of Sciences Grant No.
XDB09000000 and NSFC grant 11333003 (SM).
This research has made use of \xmm archival data.
\xmm is an ESA science mission with instruments
and contributions directly funded by ESA Member States and the USA (NASA).
In addition, the research used observations obtained with the \cha X-ray observatory,
which is operated by Smithsonian Astronomical Observatory on behalf of NASA.

\end{acknowledgements}

\end{document}